\documentstyle[psfig]{l-aa} 

\begin{document}

\thesaurus{02        
           (11.03.4 A1300; 
            11.03.1 ; 
            12.03.3 ; 
            12.04.1 ; 
            13.25.2)} 

\title{X-ray/Optical analysis of the cluster of galaxies 
Abell 1300: \\ Indications of a post-merger at z = 0.31}

\author{L. L\'emonon\inst{1} \and M. Pierre\inst{1} \and
R. Hunstead\inst{2} \and A. Reid\inst{2} \and Y. Mellier\inst{3,4}
\and H. B\"ohringer\inst{5} }

\offprints{ L. L\'emonon \small llemonon@cea.fr}

\institute{CEA Saclay DSM/DAPNIA/SAp, Service d'Astrophysique, F-91191
Gif sur Yvette \and
School of Physics, The University of Sydney, NSW 2006, Australia 
\and Institut d'Astrophysique de Paris,
98bis Boulevard Arago, F-75014, Paris
\and Observatoire de Paris, DEMIRM, 61 avenue de l'Observatoire, 
F-75014 Paris
\and Max-Planck-Institut
f\"ur Extraterrestrische Physik, D-85740 Garching}

\date{Received ??? , Accepted ???}  

\maketitle

\markboth{L\'emonon et al.: X-ray/optical analysis of A1300}{L\'emonon et al.: X-ray/optical analysis of A1300}

\begin{abstract}

We present ROSAT PSPC and HRI observations of the distant cluster
A1300 ($z = 0.3071$) in conjunction with optical data, to investigate
the physics of the intra-cluster medium and the dynamical state of the
cluster. By means of a multi-resolution wavelet analysis, we find
evidence for structures in the X-ray emission of this very luminous
($L_{X} \sim 1.7 \times 10^{45}$~erg\,s$^{-1}$
%
in the 0.1--2.4 keV band) and massive ($M_{\rm tot} \sim 1.3 \times
10^{15}$ M$_{\odot}$ at a radius $\sim$ 2.2 Mpc) cluster. This cluster
is significant as we appear to be witnessing the end of a merger which
occurred at an early epoch.

\end{abstract}

\section {Introduction}

A significant fraction of clusters of galaxies appears to be
undergoing mergers: 30\,\% of those surveyed with EINSTEIN (Forman \&
Jones, 1990), although this is certainly an underestimate as deeper
ROSAT exposures show evidence for mergers even in apparently relaxed
clusters such as Coma (White et al., 1993) or A2256 (Briel et al.,
1991). How this fraction evolves with redshift is still an open
question, but the answer is crucial as it provides important
cosmological constraints and has implications for hierarchical
clustering.

Some of the substructure we see in nearby clusters --- such as Coma
--- implies only minor disturbances related to comparatively small
mass components. The detection of substructure in distant clusters,
however, naturally implies much larger disturbances because of the
lower spatial resolution.  If this could be quantified 
(and discriminated from other environmental effects) 
it could be used as a measure of the degree of cluster growth as a
function of time, and so constrain fundamental cosmological
parameters, i.e., the amount of dark matter, index of the primordial
fluctuation power spectrum and $\Omega$. However, as observational
difficulties become greater with distance, multi-wavelength studies of
particularly luminous clusters are required to be able to discern any
evidence of a merger.

In a high resolution multi-wavelength campaign (IR, radio, optical and
X-ray) initiated in order to study in detail the properties of distant
bright X-ray clusters newly discovered in the ROSAT All-Sky Survey
(RASS) (Pierre et al., 1994a), a sample of $\sim$10 clusters having
X-ray luminosities greater than $10^{44}$ erg\,s$^{-1}$ was selected,
covering the redshift range $0.1<z<0.3$ (Pierre et al., 1994b).  The
cluster of galaxies A1300, which is described as a richness class 1
object (Abell et al., 1989), was found to be among the most luminous
($L_{X} > 10^{45}$ erg\,s$^{-1}$ in the ROSAT hard band), most
extended ($\sim 3$~Mpc) and most distant ($\overline{z} =
0.3071$) in our sample.

Subsequent detailed spectroscopy and photometry of the main body of
the cluster yielded a high velocity dispersion ($ \sigma_{v}
\sim 1200 $ km\,s$^{-1}$: Pierre et al., 1997, hereafter Paper I).
This large value is hard to reconcile with the picture of a single
well relaxed cluster. Although the histogram of the velocity
dispersion did not show significant substructure, the number density
contours indicated the presence of several sub-groups of galaxies,
especially East and North of the cluster center. These facts taken
together suggest that A1300 has undergone a merger at some stage. This
hypothesis is now investigated in detail with additional optical data
for the Northern part of the cluster combined with a study of the
X-ray morphology.  An associated radio study will follow (Reid et al.,
1997, in preparation; hereafter Paper III).

The paper is structured as follows: the optical observations are
presented in section 2, the X-ray observations in section 3, and the
merging hypothesis is then discussed in section 4 where a reference is
made to the first results of the radio study.

Throughout the paper we adopt a cosmological model with $H_0=50$
km\,s$^{-1}$Mpc$^{-1}$ and $q_{0} = 1/2$. Celestial coordinates are in
J2000.

\section {Optical data}

\subsection{Observations}

The data were collected during a three night run at the New Technology
Telescope (NTT) at La Silla in February 1996, where four clusters were
observed with the ESO Multi-Mode Instrument (EMMI). The selected mode
was Red Imaging and Low Dispersion spectroscopy (RILD) and the CCD was
the Tek \#36 (2k $\times$ 2k pixels of 24 $\mu$m). This is a
thinned, back-illuminated CCD with a quantum efficiency of better than
60\% between 400 and 800 nm. The observing configuration provides a
pixel size of 0\farcs27 over a field of view of about $9\arcmin
\times 9\arcmin$.

\begin{figure*}
\psfig{file=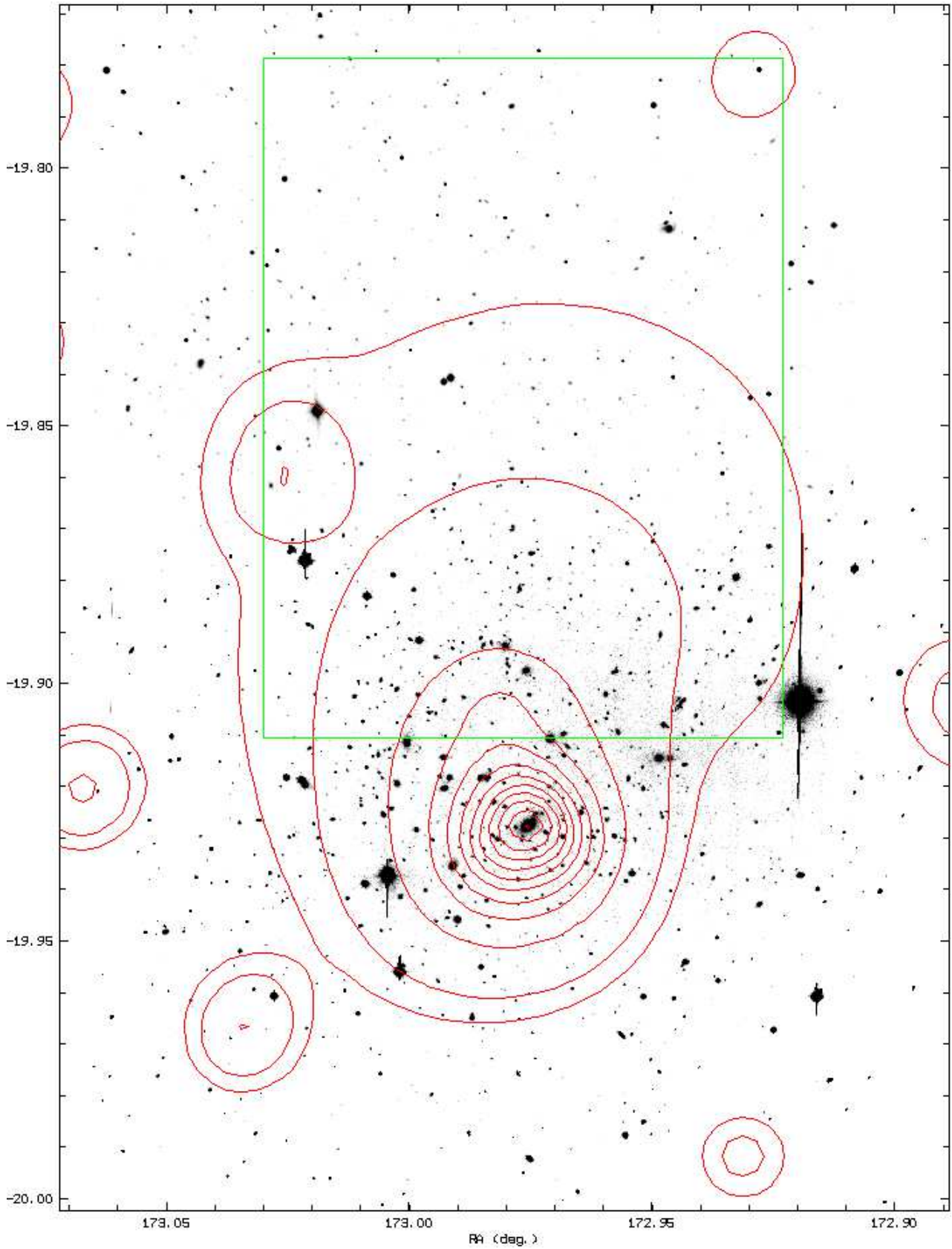,width=17cm}
\caption []{ Optical field of A1300, observed at the CFHT in R band for 
the southern part, and at the NTT in V band for the northern part.
Overlaid are the ROSAT PSPC contours in the energy range 0.4--2.4 keV
with levels [2.3, 5 (16) 165] $\times
10^{-7}$~counts\,\,s$^{-1}$\,arcsec$^{-2}$.
The first contour corresponds to a 3$\sigma$ level detection.  The
X-ray image has been filtered using a multi-resolution wavelet
analysis and pointing corrected to align the X-ray peak with the cD
galaxy (see section \ref{wavelet}). The box shows the northern area
covered by the new spectroscopic data (see Fig.\ \ref{chart}). The
displayed field is 11\arcmin\ ($\alpha$) $\times$ 15\arcmin\
($\delta$).
\label{superover} }
\end{figure*}

\begin{figure}
\psfig{file=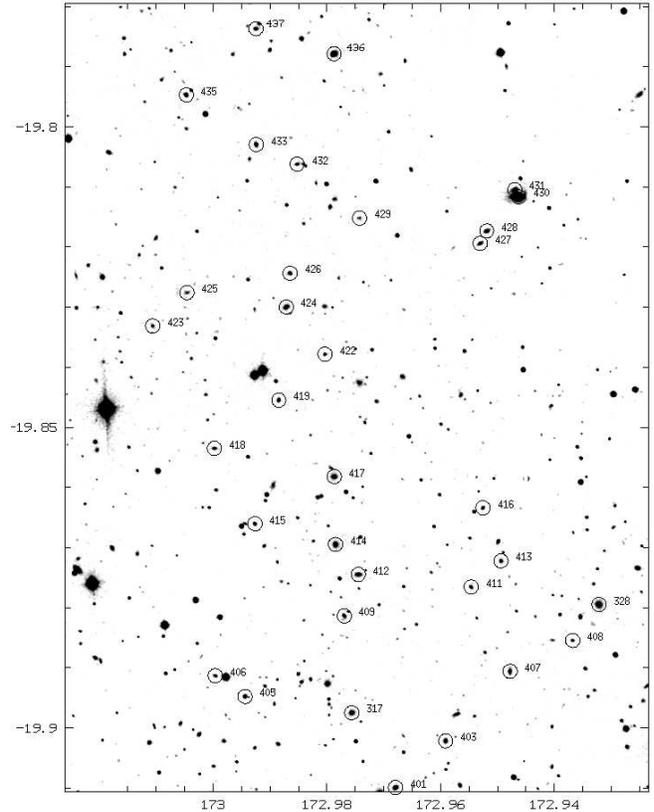,width=8.5cm}
\caption []{ Finding chart for galaxies with new measured redshifts.  
North is up and east is to the left; the field of view is 6\farcm2
($\alpha$) $\times$ 7\farcm9 ($\delta$).  This area covers the
northern extent of the hard X-ray image (see box in Fig.
\ref{superover}).  The reference numbers refer to Table
\ref{spectro}, and to Paper I for objects 317 and 328.   
\label{chart} }
\end{figure}

The seeing was good for the whole run with a stellar FWHM of
$\sim$1\farcs1.  From our previous CFHT data (Paper I) there was no
obvious fall-off in galaxy surface density towards the north, and the
X-ray emission is also extended in this direction (Fig.\
\ref{superover}), so we wanted to investigate this area in more
detail.  One mask of 37 spectra was taken for a field located some
6\arcmin\ north of the cD galaxy (Fig.~\ref{chart}).
Observations are summarized in Table \ref{observation}.

\begin{table}
\caption[ ]{Observing summary for A1300.  The cD position (J2000) is 
11h31m54.1s, $-$19$^{\circ}$55\arcmin40\arcsec.  The 1993 run yielded
64 spectra (see Paper I) and the new run, shown in bold, 37.
\label{observation}}
\begin{flushleft}
\begin{tabular}{lccc}
\noalign{\smallskip} \hline \hline \noalign{\smallskip} & Field
& Integration & Observing \\ & center & time (min) & run \\
\noalign{\smallskip} \hline
\noalign{\smallskip} {\em Photometry} \\ R images & cD & 15, 10 & CFHT 93 \\ B
images & cD & 2$\times$25 & CFHT 93 \\ {\bf V image} & {\bf 6\arcmin \/ 
north}
 & {\bf 5} & {\bf NTT 96} \\
 & {\bf of the cD} & & \\ \hline \noalign{\smallskip} {\em Spectroscopy} \\
 1st mask
& cD & 45, 40 & CFHT 93 \\ 2nd mask & cD & 75, 50 & CFHT 93 \\ 3rd mask & 
3\arcmin \/ 
north & 60, 75 & CFHT 93 \\ & of the cD & & \\ {\bf 4th mask} & {\bf 
6\arcmin \/ north} & 
{\bf 2$\times$55} 
& {\bf NTT 96} \\ & {\bf of the cD} & & \\
\noalign{\smallskip} \hline \hline
\end{tabular}
\end{flushleft}
\end{table}


\subsection{Quick photometric analysis}

In order to compare the X-ray map with the light distribution of the
cluster, a photometric catalog of the galaxies was produced from the
5-minute V-band image of the present target area.  The photometric
analysis was performed by means of the SExtractor package (Bertin \&
Arnouts 1995) in the same way as Paper I, but adapted to our data. The
V image was first cleaned of cosmic ray events (local clipping and
median filtering), then slightly smoothed by a $\sigma=1$ pixel
Gaussian (comparable with the seeing), and finally the background was
estimated using a 32 $\times$ 32 pixel mesh.  Source detections were
claimed if at least 9 adjacent pixels were above a threshold
corresponding to 3 times the local background level.  The photometric
calibration was done using the UBVRI Johnson-Kron-Cousins system
(Landolt 1992) in SA98, with the stars observed during the same run.
Estimates of the photometric errors were taken directly from the
SExtractor analysis, and are summarized in Table~\ref{errors}.

\begin{table}
\caption[ ]{Maximum and mean RMS errors in the photometric data,
 as estimated by the SExtractor program.  \label{errors}}
\begin{flushleft}
\begin{tabular}{cll}
\noalign{\smallskip} \hline \hline \noalign{\smallskip} Magnitude &
\multicolumn{1}{c}{Mean RMS error} & \multicolumn{1}{c}{Max. RMS error} \\ 
 \hline 
 V $\leq$ 20 & \hspace*{1cm}0.011 & \hspace*{1cm}0.05 \\
 20 $<$ V $\leq$ 21 & \hspace*{1cm}0.018 & \hspace*{1cm}0.065 \\
 21 $<$ V $\leq$ 22 & \hspace*{1cm}0.033 & \hspace*{1cm}0.17 \\
 22 $<$ V $\leq$ 23 & \hspace*{1cm}0.065 & \hspace*{1cm}0.29 \\
 V $>$ 23 & \hspace*{1cm}0.11 & \hspace*{1cm}0.51 \\ \noalign{\smallskip} 
\hline \hline
\end{tabular}
\end{flushleft}
\end{table}

\begin{figure}
\psfig{file=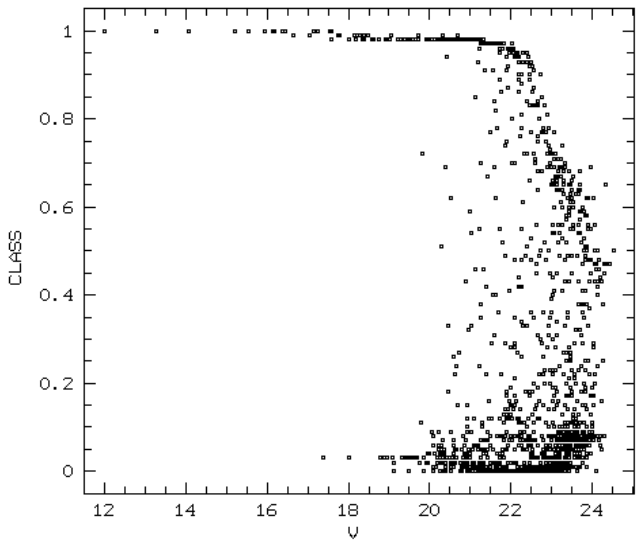,width=8.5cm,angle=270}
\caption []{ SExtractor object classification as a function of V magnitude 
for the A1300 field shown in Fig.\ \ref{chart}.  The SExtractor
star/galaxy classifier ranges from 0 (galaxy) to 1 (star).  All
objects up to class = 0.95 are expected to be galaxies, i.e. 702 out
of 832 objects up to V = 23, the magnitude completness (see text).
\label{classmag} }
\end{figure}

The catalog is estimated to be complete to V = 23. The fraction of
galactic stars was estimated using the Robin et al.\ (1995) model, and
some 130 stars with 10 $\leq$ V $\leq$ 23 are expected in our field.
This means that detected objects within this magnitude range having a
SExtractor classification $<0.95$ may be assumed to be galaxies, i.e.,
702 objects (see Figure \ref{classmag}).  Changing the threshold does
not affect the outcome significantly because most of the galaxies are
well separated from stars (3/4 of the galaxies fall below 0.2).

\begin{figure}
\psfig{file=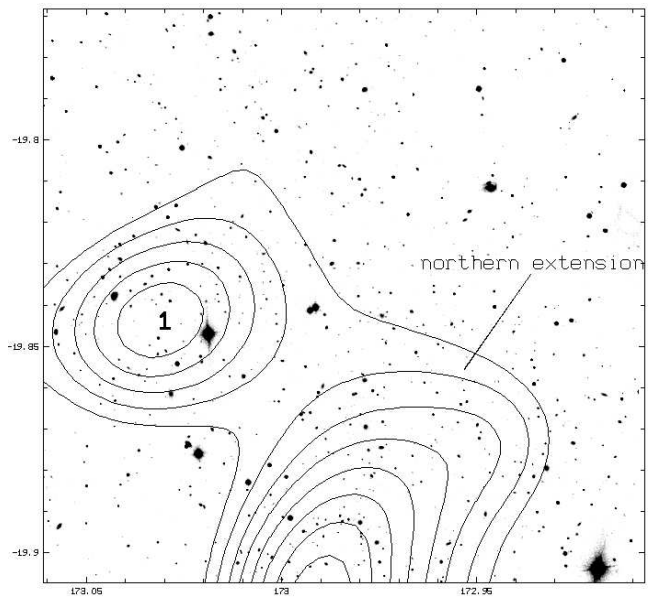,width=8.5cm,angle=270}
\caption []{ Galaxy isodensity contours for galaxies with ${\rm V} < 23$
overlaid on the CCD image of the field centered 6\arcmin \/ N of the
cD galaxy (Fig.\ \ref{chart}).  A multi-resolution wavelet analysis
was applied.  The first contour corresponds to 10
galaxies\,arcmin$^{-2}$ (3$\sigma$ level above background), and the
contour increment is 1.2 galaxies\,arcmin$^{-2}$.  A northern
extension is clearly seen in the galaxy distribution, corresponding to
the northern X-ray extension.  Moreover, a further galaxy
concentration, labelled 1, is clearly detected (see text). The image
size is 8\farcm8 ($\alpha$)\/ $\times$ 8\farcm5 ($\delta$).
\label{isonum} }
\end{figure}

\begin{figure}
\psfig{file=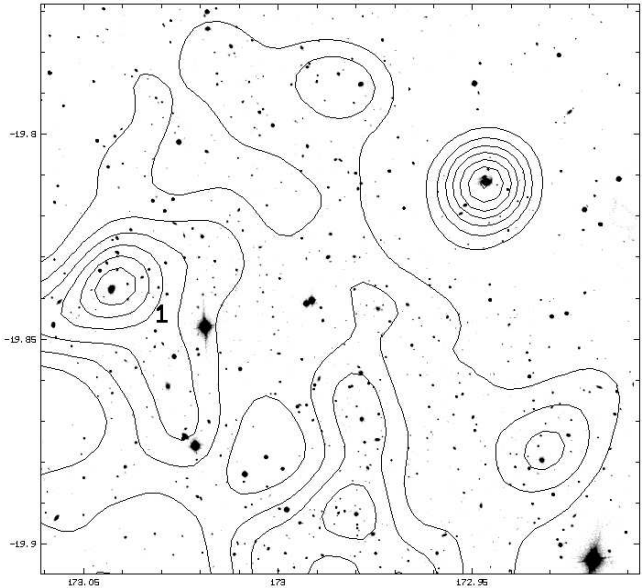,width=8.5cm,angle=270}
\caption []{ Luminosity weighted density contours for the data of 
Fig.\ \ref{isonum}. The filtering function has a $\sigma$-value of
27\arcsec. The contour increment is constant in flux, with the first
and last contours corresponding respectively to 26.5 and 25.0
mag\,arcmin$^{-2}$. The principal extended structure is 5\arcmin\ away
from the cluster center, i.e. 1.7 Mpc.  The label 1 indicates the
position of a clump detected in the galaxy density image; it is
contaminated here by a bright galaxy. The very bright point-like
structure to the northwest arises from two interacting galaxies which
do not belong to the cluster (see Table \ref{spectro}). Image size is
8\farcm8 ($\alpha$) $\times$ 8\farcm5 ($\delta$).
\label{isolight} }
\end{figure}

Using all detected galaxies brighter than V = 23, we derive two
isopleth maps, one showing galaxy number density (Fig. \ref{isonum})
and the other galaxy luminosity density (Fig.\ \ref{isolight}). These
plots cannot be compared directly with those of Paper I (where only
galaxies detected in both R and B were selected), but the overall
morphology in the overlapping region is the same.  We clearly see the
northern extension which reaches up to 5\arcmin \/ from the center,
i.e. 1.7 Mpc at the cluster redshift.  Moreover, there seems to be a
further galaxy concentration to the northeast (labelled 1 in Fig.\
\ref{isonum}).  


\subsection{Spectroscopy}

Grism \#3 was used for the spectroscopy. It has a zero deviation at
4600~\AA, covers approximately 3850--8700~\AA, and gives a dispersion
of 2.3~\AA$/$pixel. The slit has a width of 1\farcs87, i.e. 6.9
pixels, yielding a resolution of $\sim$16~\AA\ FWHM. Wavelength
calibration was done using internal Helium and Argon lamps and
subsequent reduction was performed as described in Paper I.  Redshifts
were measured by a cross-correlation method following Tonry and Davis
(1979) and implemented in the MIDAS environment. Of the 37 spectra,
two were chosen to be in common with those of the previous CFHT run to
check internal consistency. Objects 317 and 328, which were previously
found to have redshifts of $z = 0.3076$ and $z = 0.2575$ respectively
(see Paper I), are in very good agreement ($z = 0.3080$ and $z =
0.2578$); the mean difference of $\sim$ 80 km\,s$^{-1}$ is well within
the quoted spectroscopic uncertainties.

The results from the cross-correlation analysis for faint spectra were
checked by hand and the presence of a few conspicuous absorption lines
were required for them to be included in the final list which is
presented in Table~\ref{spectro}.  Heliocentric correction has not
been applied, but is negligible at this resolution.

\begin{table*}
\caption[ ]{ Spectroscopic analysis of the northern part of A1300:\\[3mm]

\begin{minipage}{11cm}

Column 1: internal reference number to Fig. \ref{chart} (and Paper I
for galaxies 317 and 328). \\[-2mm]

Column 2 \& 3: RA and Dec (J2000), decimal degrees. Galaxy positions
were determined from the V image and should have an accuracy of $\sim$
0\farcs7 rms.\\[-2mm]

Column 4: redshift\\  (*) signifies the presence of emission lines: \\
\hspace*{5mm}401: [O\,{\sc ii}], [O\,{\sc iii}], H$\beta$
, H$\alpha$, [N\,{\sc ii}], [S\,{\sc ii}] \\
\hspace*{5mm}405: [O\,{\sc ii}], [O\,{\sc iii}], H$\gamma$, H$\beta$ \\
\hspace*{5mm}435: [O\,{\sc ii}], [O\,{\sc iii}], H$\beta$ \\
$(\dagger)$: two interacting galaxies. No emission lines. \\[-2mm]

Column 5: $\Delta z$ is the internal measurement error and is related
to the correlation coefficient $c$ by the formula $\Delta z =
k/(1+c)$ where $k \sim 150$ km\,s$^{-1}$ was determined by the Tonry and
Davis (1979) method. Comparison with redshifts for galaxies in common
with Paper I suggests that $\Delta z$ is conservative. \\[-2mm]

Column 6: redshift measurement quality:\\
\hspace*{5mm}1: highest peak in the correlation function, \\
\hspace*{5mm}2: interactive measurement using lines, \\
\hspace*{5mm}9: spectrum too noisy to be verified by hand. \\[-2mm]

Column 7: V magnitude: \\
\hspace*{5mm}(1) indicates the presence of neighbors, bright and close 
enough to bias magnitude determination significantly ($\Delta$V $\sim$ 0.1). \\
\hspace*{5mm}(2) indicates that the object was originally blended with 
another one. \\[-2mm]

Column 8: Cluster member galaxy (3$\sigma$ clipping method). \\
\end{minipage}
 \label{spectro}}
\begin{flushleft}
\begin{tabular}{cccl@{\hspace{6mm}}lclc}
\noalign{\smallskip} \hline \hline \noalign{\smallskip} ID &
RA(J2000) & Dec(J2000) & \multicolumn{1}{c}{$z$} & 
\multicolumn{1}{c}{$\Delta z$ } & Q & \multicolumn{1}{c}{V} & member \\
\hline \noalign{\smallskip}
401 & 172.9679  &  $-$19.9099  & 0.2413\rlap{$^{(*)}$} & 0.0010 & 2 & 19.63 &  N \\
403 & 172.9591  &  $-$19.9022  & 0.0029 & 0.0011 & 9 & 20.28 &  N \\
317 & 172.9756  &  $-$19.8975  & 0.3080 & 0.0007 & 1 & 19.48\rlap{$^{(2)}$} &  Y \\
405 & 172.9944  &  $-$19.8948  & 0.2955\rlap{$^{(*)}$} & 0.0012 & 2 & 20.10 &  Y \\
406 & 172.9996  &  $-$19.8913  & 0.3098 & 0.0010 & 1 & 20.61 &  Y \\
407 & 172.9478  &  $-$19.8906  & 0.2572 & 0.0012 & 9 & 20.14 &  N \\
408 & 172.9368  &  $-$19.8855  & 0.3696 & 0.0012 & 9 & 21.24 &  N \\
409 & 172.9770  &  $-$19.8814  & 0.2710 & 0.0017 & 2 & 20.31\rlap{$^{(1)}$} &  N \\
328 & 172.9322  &  $-$19.8795  & 0.2578 & 0.0007 & 1 & 18.95 &  N \\
411 & 172.9547  &  $-$19.8766  & 0.3079 & 0.0010 & 1 & 20.53 &  Y \\
412 & 172.9745  &  $-$19.8745  & 0.2548 & 0.0011 & 1 & 19.94\rlap{$^{(1)}$} &  N \\
413 & 172.9494  &  $-$19.8722  & 0.2560 & 0.0009 & 1 & 20.64 &  N \\
414 & 172.9785  &  $-$19.8695  & 0.3054 & 0.0007 & 1 & 19.45 &  Y \\
415 & 172.9927  &  $-$19.8660  & 0.3075 & 0.0010 & 1 & 20.66 &  Y \\
416 & 172.9526  &  $-$19.8634  & 0.3111 & 0.0009 & 1 & 20.72 &  Y \\
417 & 172.9787  &  $-$19.8582  & 0.3013 & 0.0010 & 1 & 19.52\rlap{$^{(1,2)}$} &  Y \\
418 & 172.9998  &  $-$19.8535  & 0.2773 & 0.0011 & 1 & 20.63 &  N \\
419 & 172.9885  &  $-$19.8455  & 0.1727 & 0.0009 & 9 & 20.44 &  N \\
422 & 172.9804  &  $-$19.8378  & 0.3070 & 0.0012 & 1 & 21.00 &  Y \\
423 & 173.0107  &  $-$19.8331  & 0.2158 & 0.0010 & 9 & 20.93\rlap{$^{(1)}$} &  N \\
424 & 172.9872  &  $-$19.8300  & 0.3074 & 0.0009 & 1 & 19.57\rlap{$^{(2)}$} &  Y \\
425 & 173.0046  &  $-$19.8276  & 0.1851 & 0.0011 & 9 & 20.80\rlap{$^{(1,2)}$} &  N \\
426 & 172.9865  &  $-$19.8244  & 0.2777 & 0.0010 & 9 & 20.58 &  N \\
427 & 172.9531  &  $-$19.8194  & 0.1606 & 0.0012 & 9 & 20.53 &  N \\
428 & 172.9519  &  $-$19.8173  & 0.0482 & 0.0011 & 9 & 20.27 &  N \\
429 & 172.9743  &  $-$19.8152  & 0.3878 & 0.0011 & 9 & 21.26 &  N \\
430 & 172.9463  &  $-$19.8117  & 0.1578\rlap{$^{(\dagger)}$} & 0.0006 & 1 & 17.38\rlap{$^{(2)}$} &  N \\
431 & 172.9470  &  $-$19.8105  & 0.1592\rlap{$^{(\dagger)}$} & 0.0006 & 2 & 20.07\rlap{$^{(2)}$} &  N \\
432 & 172.9853  &  $-$19.8062  & 0.3683 & 0.0010 & 1 & 20.85\rlap{$^{(1,2)}$} &  N \\
433 & 172.9924  &  $-$19.8030  & 0.2311 & 0.0007 & 1 & 20.06 &  N \\
435 & 173.0047  &  $-$19.7947  & 0.1519\rlap{$^{(*)}$} & 0.0009 & 2 & 20.26\rlap{$^{(1)}$} &  N \\
436 & 172.9788  &  $-$19.7879  & 0.2311 & 0.0006 & 1 & 18.88 &  N \\
437 & 172.9925  &  $-$19.7837  & 0.2294 & 0.0007 & 1 & 20.27\rlap{$^{(1,2)}$} &  N \\
\noalign{\smallskip} \hline \hline
\end{tabular}
\end{flushleft}
\end{table*}

\begin{figure} 
\psfig{file=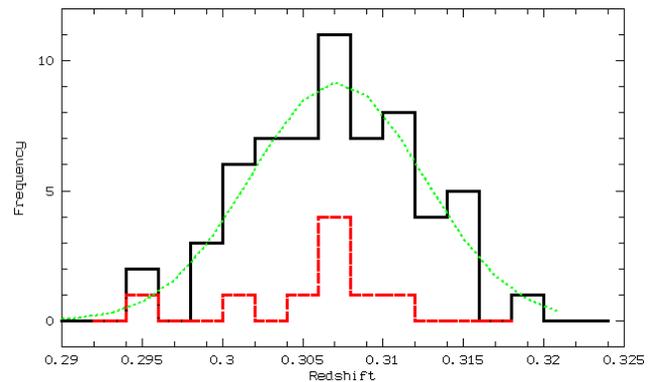,width=8.5cm,angle=270}
\caption []{ Cluster galaxy redshift histogram (bin size $\Delta z =
0.002$); the mean redshift is 0.3071 and the velocity dispersion $\sim$
1200 km\,s$^{-1}$. The solid line represents the whole dataset (Paper
I + present data) while the dashed line shows the nine new spectra.
Dotted line is the best Gaussian fit.
\label{histogram} }
\end{figure}

Cluster membership was assessed using a 3$\sigma$ clipping method and
10 of the 37 galaxies were found to belong to the cluster (one of them
in common with Paper I). The northernmost cluster galaxy (\#424) is
5\farcm9 (i.e. $\sim$2 Mpc at the cluster redshift) from the cD
galaxy. The cluster redshift distribution is displayed in Fig.
\ref{histogram}. As already noted in Paper I, while the histogram is
very broad, it does not show any significant substructure. Indeed,
the whole dataset is well fitted by a Gaussian centered on $z =
0.3071$ with $\sigma = 0.0052 \pm 0.0005$, giving a radial velocity
dispersion of $\sim 1200$ km\,s$^{-1}$.

The velocity distribution was investigated in more detail using the ROSTAT
package (Beers et al., 1990; Bird \& Beers, 1993). Skewness and kurtosis
($-$0.092 and 2.683 respectively) strongly suggest a Gaussian
distribution (formally: 0 and 3 respectively). 
We have also investigated the presence of substructures in ($\alpha$,
$\delta$, $z$) space by producing a velocity map. No obvious signal
was detected although more redshifts are required for a proper
statistical analysis.

\section{X-ray data}

\subsection{Observations}

Pointed ROSAT Position Sensitive Proportional Counter (PSPC)
observations were obtained in the period June 27--28 1993, with four
exposures totalling 8572 sec. The PSPC has an on-axis spatial
resolution of $\sim$ 30\arcsec \/ (FWHM), is sensitive to photons with
energies between 0.1 and 2.4 keV, and has an energy resolution of
about 40\% at 1 keV (Briel et al., 1988).  Data were analyzed using
the EXSAS package in MIDAS.

\begin{figure}
\psfig{file=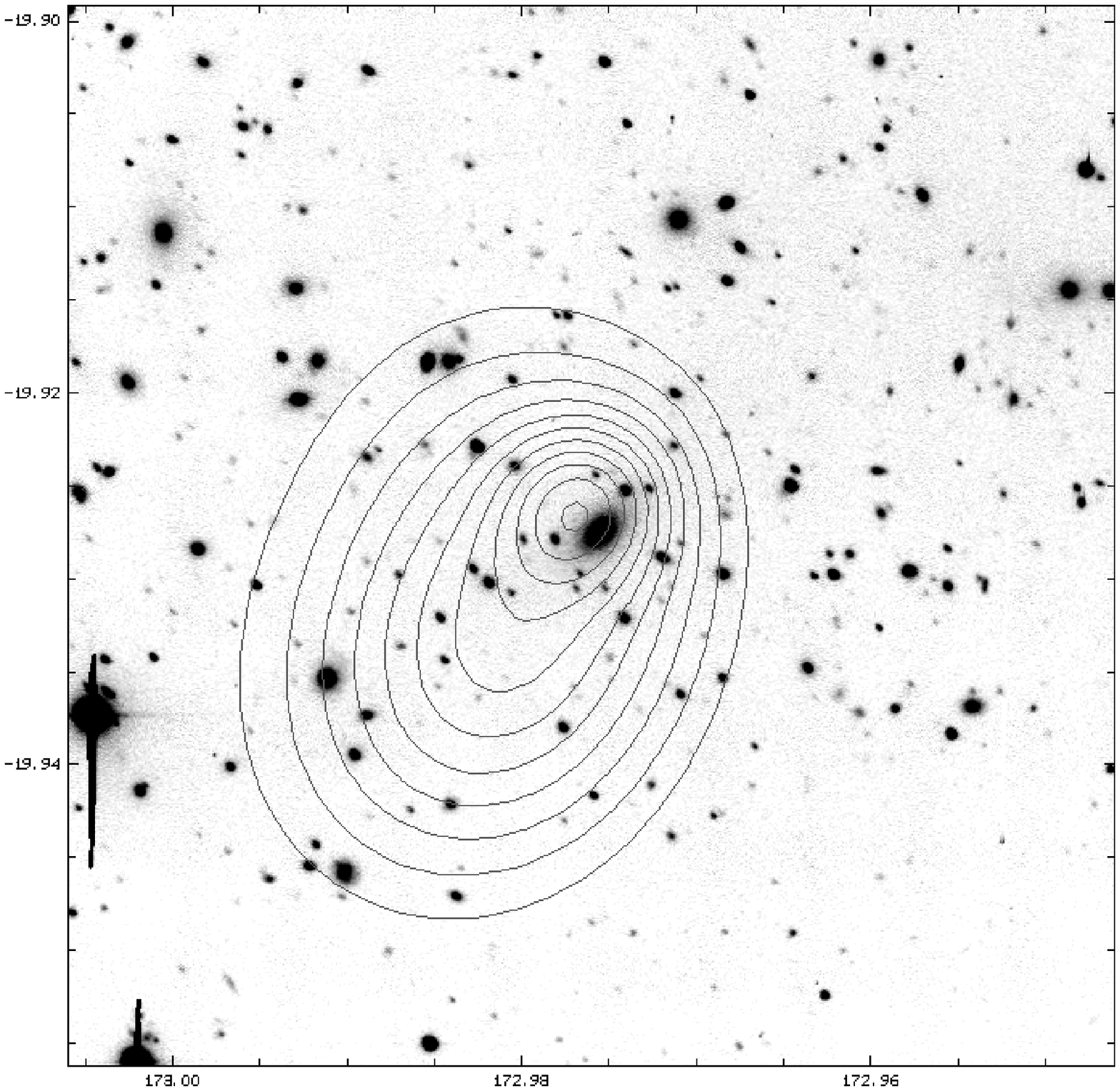,width=8.5cm,angle=270}
\caption []{ROSAT HRI image of the central region of A1300 
($t_{\rm exp} = 10^{4}$ s, $\sim$200 excess photons above background)
overlaid on the CCD image.  The contours are equally spaced from 3--8
$\times 10^{-6}$ counts\,s$^{-1}$\,arcsec$^{-2}$.  The offset between
the X-ray peak and the cD galaxy of $\sim$5\arcsec\, is assumed to
be due to satellite attitude error.  Image size is 3\farcm6 $\times$
3\farcm4.
\label{hri} }
\end{figure}

In addition, a 11,000\,s High Resolution Imager (HRI) observation of
the central region of A1300 was obtained on July 9 1994 (3,000\,s) and on
June 14--15 1995 (8,000\,s).  Exposure time
was inadequate to reach the desired signal-to-noise ratio but was
sufficient to show an elongated structure which peaks on the cD galaxy
and extends to the south east.  The contour image is shown in Figure
\ref{hri} overlaid on the R-band CCD image.

The following sections will concentrate on the analysis of the PSPC image.

\subsection{Source detection}

A maximum likelihood search for discrete sources was carried out
over the PSPC 0.1--2.4 keV image and the results are listed in Table
\ref{source}.  The X-ray point source positions were correlated with
the SIMBAD\footnote{http://simbad.u-strasbg.fr/Simbad} and
NED\footnote{telnet://ned@ned.ipac.caltech.edu/} on-line catalogs
but no identifications were found. Comparison with the
COSMOS\footnote{ http://xweb.nrl.navy.mil/www\_rsearch/RS\_form.html}
source list, derived from a circular region of radius 20\arcsec\
centered on each object from Table~\ref{source}, produced a number of
possible optical identifications.  These are listed in
Table~\ref{cosmos}.

\begin{table}
\caption[ ]{ Discrete X-ray sources in the 45\arcmin\ diameter field 
of A1300 detected with a maximum likelihood greater than 10.
Coordinates refer to the normal attitude solution, which is
likely to have a 9\arcsec\ east and 9\arcsec\ south shift with
respect to absolute celestial coordinate (see text). 
\label{source}}
\begin{flushleft}
\begin{tabular}{ccccc}
\noalign{\smallskip} \hline \hline \noalign{\smallskip} ID &
RA(J2000) & Dec(J2000) & $f_{\rm X}$ & error \\
 & h~~m~~s & ~~$^{\circ}$ ~~\arcmin\ ~~\arcsec & 
\multicolumn{2}{c}{counts~s$^{-1}$}\\
\hline \noalign{\smallskip} 
 1  & 11 31 14.8  & $-$19 45 17  & 0.0027 & 0.0009 \\
 2  & 11 31 48.7  & $-$19 50 43  & 0.0026 & 0.0008\\ 
 3  & 11 32 06.3  & $-$19 51 35  & 0.0047 & 0.0010\\ 
 4  & 11 31 52.5  & $-$19 52 01  & 0.0042 & 0.0010\\ 
 5  & 11 32 33.9  & $-$19 52 47  & 0.0031 & 0.0008\\ 
 6  & 11 31 58.3  & $-$19 52 57  & 0.0121 & 0.0016\\ 
 7  & 11 31 45.0  & $-$19 53 48  & 0.0081 & 0.0014\\ 
 8  & 11 31 54.6  & $-$19 54 17  & 0.0410 & 0.0025\\ 
 9  & 11 31 48.2  & $-$19 54 51  & 0.0089 & 0.0014\\ 
10  & 11 31 12.2  & $-$19 55 08  & 0.0042 & 0.0010\\ 
11  & 11 32 16.3  & $-$19 55 20  & 0.0050 & 0.0010\\ 
12  & 11 31 54.8  & $-$19 56 02  & 0.0896 & 0.0035\\
13  & 11 31 47.1  & $-$19 56 11  & 0.0048 & 0.0011\\
14  & 11 32 08.4  & $-$19 58 07  & 0.0043 & 0.0009\\
15  & 11 31 28.5  & $-$19 59 00  & 0.0031 & 0.0008\\
16  & 11 30 54.7  & $-$20 12 10  & 0.0056 & 0.0014\\ 
\noalign{\smallskip} \hline \hline
\end{tabular}
\end{flushleft}
\end{table}

\begin{table}
\caption[ ]{ Possible COSMOS optical identifications for the
discrete X-ray sources detected in the ROSAT field. As some sources
fall outside the CCD field, we adopt the COSMOS magnitude
for uniformity. Column 1 refers to the X-ray ID, columns 2 \& 3 give
the optical position of potential optical counterparts within a
20\arcsec\ radius, column 4 gives the ROSAT-optical offset in arcsec,
column 5 gives the COSMOS $b_{\rm J}$ magnitude, and column 6 gives
the COSMOS classification: s = star, g = galaxy, f = faint source.
\label{cosmos}}
\begin{flushleft}
\begin{tabular}{cccrcc}
\noalign{\smallskip} \hline \hline \noalign{\smallskip} ID &
RA(J2000) & Dec(J2000) & \multicolumn{1}{c}{$\Delta r$} & $b_J$ & class \\
 & $\!\!$h~~m~~~s & ~~$^{\circ}$ ~~\arcmin\ ~~\arcsec & \multicolumn{1}{c}{\arcsec} \\
\hline \noalign{\smallskip} 
 1  & 11 31 14.28  & $-$19 45 19.0 & 7.8  & 20.0 & s\\
    & 11 31 16.07  & $-$19 45 15.4 & 18.0 & 16.7 & s  \\
 2  & 11 31 49.54  & $-$19 50 39.8 & 12.6 & 22.2 & f \\ 
 3  & 11 32 06.81  & $-$19 51 41.2 & 9.6  & 20.5 & g \\ 
    & 11 32 06.45  & $-$19 51 15.2 & 19.8 & 17.7 & s \\
 4  & 11 31 53.79  & $-$19 52 05.7 & 18.6 & 21.5 & s \\ 
 5  & 11 32 34.61  & $-$19 52 45.3 & 10.2 & 18.0 & s \\ 
 6  & 11 31 59.72  & $-$19 52 53.8 & 20.4 & 18.1 & s \\ 
 7  & 11 31 45.69  & $-$19 53 59.3 & 15.0 & 22.6 & f \\ 
    & 11 31 43.89  & $-$19 53 36.5 & 19.2 & 23.0 & f \\
 8  & 11 31 53.98  & $-$19 54 07.7 & 13.0 & 20.4 & s \\
 9  & 11 31 47.38  & $-$19 54 52.8 & 12.0 & 19.3 & s \\
    & 11 31 47.58  & $-$19 54 52.7 & 9.0  & 20.6 & g \\
    & 11 31 47.00  & $-$19 54 53.1 & 16.8 & 21.3 & g \\
11  & 11 32 16.16  & $-$19 55 20.9 & 2.4  & 20.0 & s \\ 
    & 11 32 15.33  & $-$19 55 15.5 & 14.4 & 20.1 & g \\
12  & 11 31 53.69  & $-$19 55 56.7 & 16.2 & 22.2 & f \\
    & 11 31 55.63  & $-$19 55 50.2 & 16.8 & 22.8 & f \\
    & 11 31 54.70  & $-$19 55 41.8 & 20.4 & 22.2 & f \\
14  & 11 32 08.78  & $-$19 58 09.7 & 6.0  & 22.2 & f \\
15  & 11 31 28.31  & $-$19 59 02.9 & 4.2  & 19.0 &s \\
\noalign{\smallskip} \hline \hline
\end{tabular}
\end{flushleft}
\end{table}

\subsection{X-ray morphology \label{wavelet}}

A fundamental aspect of morphological studies of the intracluster
medium (ICM) is the search for substructure. Because of its shorter
relaxation time, the X-ray gas is expected to trace the shape of the
gravitational potential more closely than galaxies.  The X-ray
morphology, therefore, is better suited to the investigation of
substructures and merging events as well as providing insights into
the dark matter distribution.

Because of the very high temperature of the ICM (10$^7$--10$^8$ K,
Sarazin 1986), the energy is mostly emitted in the range 1--10 keV.
Moreover, the soft ROSAT band, i.e. $\leq$ 0.4 keV, is strongly
contaminated by background emission, predominantly from the Galaxy.
Consequently, the photon file was binned into separate 5\arcsec\ pixel
images, with energies in the range 0.4--2.4 keV (expected to belong to
the cluster) and 0.1--0.4 keV. These images were then filtered using a
multi-resolution wavelet analysis program (Starck et al., 1996).  This
method suppresses photon noise and restores structures on different
scales, in this case with a confidence level of 99.9\%.  Figure
\ref{superover} shows the hard X-ray contours overlaid on the optical
image of the cluster.

Alignment of the optical images was obtained by computing the
transformation parameters between CCD coordinates and celestial
coordinates using ten stars with equatorial coordinates measured on
the COSMOS UKST Southern Sky Catalog. The standard deviation of the
residuals was less than 0\farcs7. Because of the small number of
photons in all of the point-like X-ray sources that were detected
(about twenty for the brightest) we were unable to check the attitude
solution of the X-ray image. We therefore decided to align the peak in
the hard band X-ray image with the optical cD galaxy. This corresponds
to a shift in the X-ray center of 9\arcsec\ west and 9\arcsec\ north.
This choice appears to be compatible with the location of some of the
point sources in the field with faint counterparts in the digitized
sky survey\footnote{http://skview.gsfc.nasa.gov/skyview.html}. Further
support for the above registration came after multi-resolution wavelet
filtering which showed that the very soft source (number 8 in Table
\ref{source} and the best spatially determined) now coincides with
the stellar object (possibly a QSO) identified in Table~\ref{cosmos}.

\begin{figure}
\psfig{file=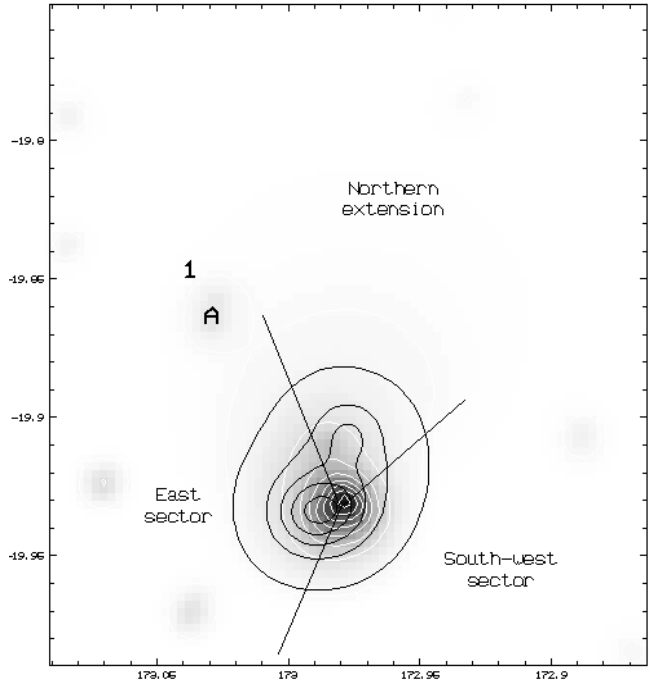,width=8.5cm}
\caption []{ROSAT PSPC images of A1300.
The soft image (0.1--0.4 keV), shown in equally spaced black solid
contours from 0.35--$1.75 \times 10^{-6}$
counts\,s$^{-1}$\,arcsec$^{-2}$, is overlaid on the hard image
(0.4--2.4 keV), shown in greyscale with equally spaced white contours
from 0.47--$16.3 \times 10^{-6}$ counts\,s$^{-1}$\,arcsec$^{-2}$.
Filtered images were obtained by a multi-resolution wavelet analysis
(see text). Label ``1'' marks the position of a local clump in the
galaxy isodensity image (Fig.\ \ref{isonum}), while ``A'' indicates a
second luminosity maximum in the hard image.  In each band the first
contour level is 3$\sigma$ above the background level.  The south-west
and east sectors refer to the spectral analysis.  Image size is
13\arcmin\ $\times$ 14\farcm5.
\label{Xmorph} }
\end{figure}

Contours of the filtered soft- and hard-band images of A1300 are
overlaid in Figure \ref{Xmorph}.  The hard-band image is clearly
asymmetric and has a higher peak count rate.  There is also a
significant displacement of $\sim$23\arcsec\ between the soft and hard
maxima.  The soft source mentioned above (no.\ 8 in Table
\ref{source}) can be seen immediately north of the hard X-ray peak.
Interestingly, we detect an additional maximum in the hard band to the
north-east. The multi-resolution analysis shows that this emission
region (labelled ``A'' in Fig.~\ref{Xmorph}) is not point-like and is
slightly elongated north-south. It falls close in projection to the
position of clump 1, detected in the galaxy isodensity image (Fig.\
\ref{isonum}); the offset of 1\farcm7 corresponds to $\sim$0.5 Mpc at
the cluster redshift.

\begin{figure}
\psfig{file=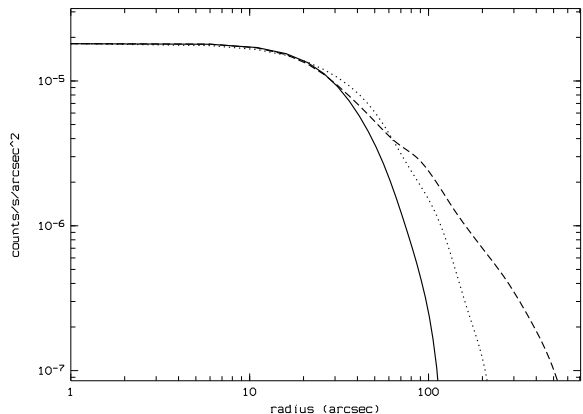,width=8.5cm,angle=270}
\caption []{ The X-ray surface brightness asymmetry: profiles 
to the east (dotted), north (dashed) and west (solid) of the peak of
the filtered hard image. The northern profile is clearly more
elongated and falls off more gently than the east and west profiles.
The steep fall off at low intensity is due to the wavelet filtering
process.  The background was estimated to be $8.5\times 10^{-8}$
counts\,s$^{-1}$\,arcsec$^{-2}$, so profiles were cut at this value.
All structures are detected with a 99.9\% confidence level.
\label{profileX} }
\end{figure}

On larger scales, the cluster shows a strong elongation to the north,
with extended emission being detected as far as 8\farcm3 north of
the the peak ($\sim$ 2.8 Mpc at this redshift).  In contrast the
western extent is only 1\farcm7 (0.5 Mpc).  This important
difference is displayed in Fig.~\ref{profileX} where the X-ray surface
brightness profiles in three different directions are compared.

\subsection{Surface brightness}

Despite its clearly asymmetrical morphology, we did not have enough
photons to further characterize the different sub-clumps. It is,
however, possible to calculate a circularly averaged X-ray surface
brightness profile of A1300 from the hard band. We removed areas
corresponding to the northern extension and all point-like sources
from the hard image (restricted to 0.5--2.0 keV) and further assumed
circular symmetry. We noticed that the profile fit was sensitive to a
shift in the X-ray center, so we smoothed the data by binning into
30\arcsec\ pixels.  The profile was then fitted with an isothermal
``$\beta$--model'' (King, 1962; Cavaliere \& Fusco-Femiano, 1976,
1981) of the form

\begin{equation}
S(R)=S_{0}[1+(R/R_{c})^{2}]^{0.5-3\beta} + {\rm background, }
 \label{king}
\end{equation}

\noindent simultaneously deconvolving by the PSPC PSF at 1 keV
(Hasinger et al., 1992). 

\begin{figure}
\psfig{file=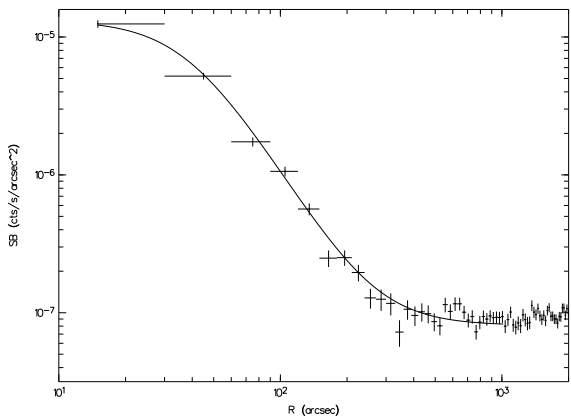,width=8.5cm,angle=270}
\caption []{ X-ray surface brightness profile of A1300 (circularly 
averaged) in the energy range 0.5--2.0 keV (channels 52--201).  A bin
size of 30\arcsec\ was used to obtain a good signal-to-noise ratio.
The solid line is the best fit to a King profile (Eq.\ \ref{king})
convolved with the PSF and overlaid on the data.  \label{profile} }
\end{figure}

The fit, convolved with the PSPC resolution, is overlaid on the data
in Fig.~\ref{profile}. The values of the fitted parameters were:
$S_{0} = (1.44 \pm 0.2) \times 10^{-5}$ counts\,s$^{-1}$\,arcsec$^{-2}$
(at a 95\% confidence level), $\beta = 0.64 \pm 0.03$, $R_{c} = 43 \pm
5$\arcsec\ ($\sim$240 kpc at the cluster redshift; both errors given
at a 99\% confidence level), and a background value of $8.1 \times
10^{-8}$ counts\,s$^{-1}$\,arcsec$^{-2}$.  Note that $\beta$ and the
core radius are only representative of the circularly averaged surface
brightness.  Their values are very close to average values for a large
number of clusters studied by Jones \& Forman (1984).  The fact that
$\beta$ is less than unity indicates that the energy per unit mass in
the gas is higher than in the galaxies and that the gas density falls
off less rapidly with radius than the galaxy density.  However, as
Fig.~\ref{profileX} shows, there are significant differences in the
profiles in different directions.

If we continue to assume circular symmetry and use the averaged
characteristics for the cluster, the surface brightness (Eq.\
\ref{king}) can be deprojected analytically.  Furthermore, the
observed surface brightness is weakly dependent on the gas temperature
(this is justified because of the expected high temperature of this
very bright cluster and the poor temperature sensitivity of the ROSAT
PSPC above 5 keV).  In this case, the proton density distribution
$n_{p}(r)$ is given by

\begin{equation}
n_{p}(r) = n_{p}(0)[1+(r/R_{c})^{2}]^{-\frac{3\beta}{2}}. 
\label{gaz3d}
\end{equation}

\noindent Using the above values for $\beta$ and $R_{c}$, and 
normalizing the density profile according to the observed count-rates,
we obtain $n_{p}(0) = 8.5\times 10^{-3}$~cm$^{-3}$. This density
will be adopted hereafter.

\subsection{Spectral analysis \label{sectspec}}

\begin{figure}
\psfig{file=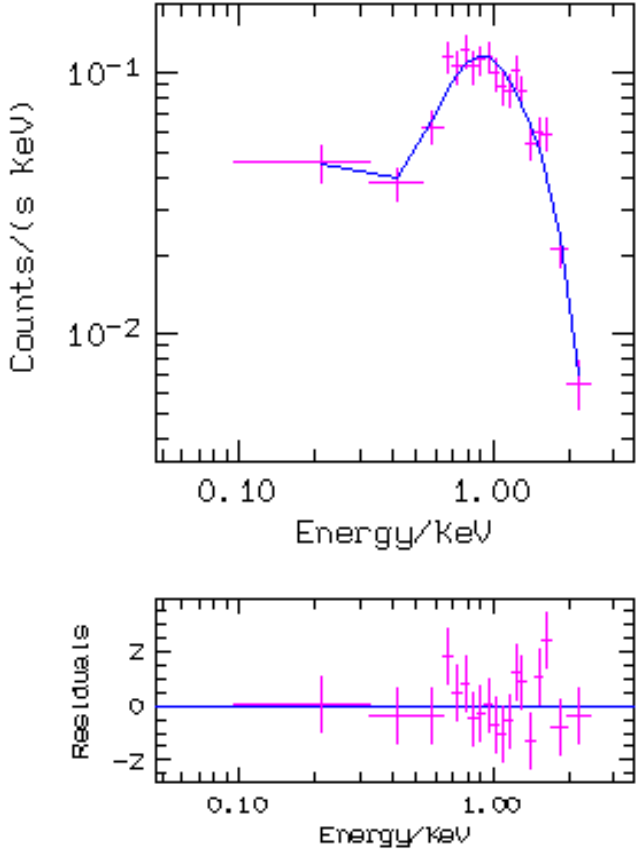,width=8.5cm}
\caption []{ Best fit of a Raymond \& Smith thermal plasma model.
 The fit indicates a temperature of $\sim 5\pm 3$~keV.
\label{spec} }
\end{figure}

\begin{figure}
\psfig{file=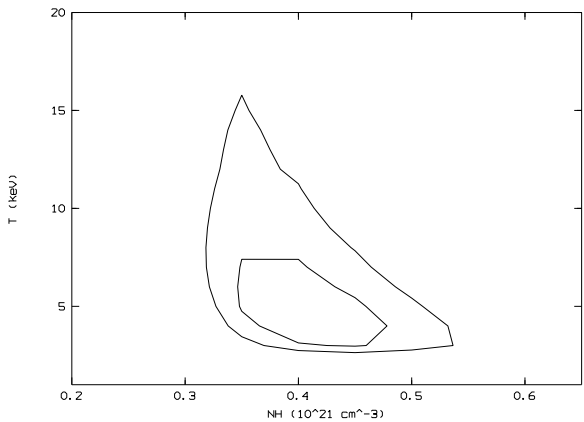,width=8.5cm,angle=270}
\caption []{ Contours of $\chi^2$ as a function of column density
$N_{H}$ and temperature T for confidence levels of 68.3\% and
95.4\%
\label{nh_t} }
\end{figure}

The ROSAT PSPC spectral range (0.1--2.4 keV) is not well suited to the
temperature measurement of bright clusters. However, we fitted data
between 0.1 and 2.4 keV with a Raymond \& Smith (1977) thermal plasma
model. The spectrum was binned such that each bin had a
signal-to-noise ratio of 7, and the bins extended out to a radius of
150\arcsec\ (850 kpc) from the center; a sector containing the soft
point-like source was excluded.  We fixed the metallicity at Z = 0.3
(variations in the abundance did not affect the results by more than
10\%) and obtained a Galactic neutral hydrogen column density of
$N_{H}=(4.1\pm 0.9)\times 10^{20}$~cm$^{-2}$, in good agreement with the
value from Dickey \& Lockman (1990) of $N_{H}=4.5\times 10^{20}$
cm$^{-2}$.  With these constraints we obtained a temperature for the
cluster of $5\pm 3$ keV ($\chi^{2} = 18$ with 16 variables).  The fit
is displayed on the spectrum in Fig.\ \ref{spec} and as $\chi^2$
contours in Fig.\ \ref{nh_t}).

The unusual morphology of A1300 described above suggested that it may
be worthwhile looking for spatially separated temperature components.
We considered (i) the south-west sector (pa 160$^\circ$--315$^\circ$,
and (ii) the east sector (pa 20$^\circ$--160$^\circ$) (cf.
Figs.~\ref{Xmorph}, \ref{profileX}).  Despite the small number of
photons ($\sim$ 500 and $\sim$ 600 respectively), we find some
indication of a temperature gradient.  As before, spectra were binned
so that each bin had a signal-to-noise ratio of 7, and fits were
renormalized between 0.9 and 1.0 keV for comparison. In both cases, we
found a Galactic $N_H$ in good agreement with the previous value.  The
best fit for the south-western sector indicates a temperature of $3\pm
2$ keV, whereas that for the eastern sector suggests a much higher
value ($>$7 keV) (both temperatures at a 68\% confidence level), but
with very large uncertainties as expected with the ROSAT sensitivity.
The temperature deduced from the galaxy velocity dispersion
correlation ($T_{X} = 10^{-3.22}\sigma^{1.35}$; Edge \& Stewart,
1991b) is 9 keV, marginally consistent with our global determination
but beyond the ROSAT spectral range. We tentatively adopt this value
in the following sections.

\subsection{Total luminosity}

The total cluster luminosity was calculated from the fitted surface
brightness profile. We integrated the ROSAT counts out to a radius
$r_{\rm cut}$ of 400\arcsec\ (2.2 Mpc) set by the detection
limit of the circularly averaged surface brightness profile (see
Fig.~\ref{profile}).  This gives an overall count rate $C$, where 

\begin{equation}
C = \frac{\pi S_{0} R_{c}^{2}}{3/2-3\beta}\left[\left(1+r_{\rm cut}^{2}
\right)^{3/2-3\beta}-1\right],
\end{equation}

\noindent of 0.25 counts\,s$^{-1}$.  This count rate was
converted into luminosity using a Raymond \& Smith (1977) thermal
model folded with the PSPC response provided by the EXSAS package, and
assuming a thermal spectrum with $T=9$~keV, metallicity of 0.3 and a
neutral hydrogen column density of $4.5\times 10^{20}$ cm$^{-2}$. We
obtain $L_{X} = (1.7\pm 0.1)\times 10^{45}$~erg\,s$^{-1}$ (at a 95\%
confidence level) in the 0.1--2.4 keV band.  Using the
empirical temperature-luminosity relation (Henry \& Arnaud 1991, Edge
\& Stewart 1991a), this would indeed correspond to a temperature 
$\sim$ 9 keV, consistent with the estimate obtained from the galaxy
velocity dispersion correlation (Sect. \ref{sectspec}).

Although no excess X-ray emission is observed at the cluster center,
we investigated the presence of a cooling flow. An estimate of the
cooling time (Sarazin 1986) is given by

\begin{equation}
T_{\rm cool}({\rm yr}) = 8.5\times 10^{10} \left(\frac{T}{10^{8}
{\rm K}}\right)^{1/2}\left(\frac{n_{p}}{10^{-3}\, {\rm cm}^{-3}}\right)^{-1}.
\end{equation}

Using the density profile determined above and T = 9 keV, we derive
$R_{\rm cool}$ which defines the region where the cooling time is
smaller than the age of the universe at the cluster redshift.
Depending on the value of the Hubble parameter $h = H_{0}/100$,
$R_{\rm cool}$ lies in the range 0--27\arcsec\ for $h= 0.75$--0.5. For
$h > 0.75$, the cooling time is greater than the age of the universe.
Calculating the bolometric cluster luminosity ($L_{\rm cool}$)
enclosed within $R_{\rm cool}$, and assuming steady state isobaric gas
cooling from 9 keV, we derive a mass flow (following Fabian et al.,
1991) of

\begin{equation}
\dot{M}_{\rm cool} = \frac{2}{5}\frac{\mu m_{p}}{kT}L_{\rm cool} 
\sim 880 \left(\frac{L_{\rm cool}}{10^{45}}\right)\!\!
\left(\frac{T}{5\,{\rm keV}}\right)^{-1}\!\!
{\rm M}_{\odot}/{\rm yr}
\end{equation}

\noindent This yields $ \dot{M}_{\rm cool} < 400$ M$_{\odot}$/yr
for $h > 0.5$, but uncertainties are large (up to a factor $\sim 2$).

\subsection{Mass determination}

We can obtain a robust estimate of the cluster mass by removing
obvious regions of disturbance and assuming hydrostatic equilibrium
for the rest of the cluster following Schindler (1996).  We have a
pressure gradient

\begin{equation}
\frac{dP}{dr} = -n\mu m_{p}\frac{d\phi}{dr}, \label{euler}
\end{equation}

\noindent where $P$ is given by the ideal gas law, $P = nkT$. 

The mass integrated out to radius $r$ is then 

\begin{equation}
M(r) = -\frac{r^{2}kT}{G\mu
m_{p}}\left(\frac{1}{\rho_{g}}\frac{d\rho_{g}}{dr}+
\frac{1}{T}\frac{dT}{dr}\right)
\end{equation}

\noindent where $\rho_{g}$ is the gas density. Using the density profile 
(Eq.\ \ref{gaz3d}) derived above, this yields

\begin{equation}
M(r) = -\frac{r^{2}k}{G\mu m_{p}}\left(-\frac{3\beta
rT}{r^{2}+R_{c}^{2}}+\frac{dT}{dr}\right). \label{masse}
\end{equation}

Since we expect the gas density profile to be accurate, the main error
in the mass determination comes from the large uncertainty in the
temperature. If we ignore any radial temperature dependence, we obtain
a cluster mass of 

\begin{equation}
M(r) = \frac{3\beta r^{3}kT}{G\mu m_{p}(r^{2}+R_{c}^{2})}
\end{equation}

\noindent which diverges with radius. As above, we integrate 
the mass out to a radius $r_{\rm cut}$ of 2.2 Mpc and obtain a total
mass of $M_{\rm tot} \sim 1.3\times 10^{15}$~M$_{\odot}$. Equation
(\ref{gaz3d}) gives a corresponding $M_{\rm gas}/M_{\rm tot}$ ratio
of $\sim$30\%, in good agreement with previous determinations for rich
clusters (Briel et al., 1992, Henry et al., 1993).

\section{Discussion and conclusions}

For the first time in such a distant and rich cluster, high resolution
multi-wavelength observations are enabling us to gain a deeper insight
into the cluster dynamics.
 
Previous optical spectroscopy of A1300 (Paper I) showed a very broad
galaxy velocity dispersion ($\sigma_v = 1200$ km\,s$^{-1}$), hardly
compatible with the picture of a relaxed cluster.  In this paper,
ROSAT PSPC observations, in conjunction with a multi-resolution
wavelet analysis, have revealed an irregular morphology which is very
extended to the north but falls off steeply to the south-west.
Indications of non-isothermality following the brightness morphology
were also found. The estimated total cluster mass is very high
($M_{\rm tot} \sim 1.3\times 10^{15}$~M$_{\sun}$) out to a radius of
2.2 Mpc, and the gas mass represents $\sim$30\% of the total mass.

A1300 is probably not in an equilibrium state as it shows signatures
of a merger in the optical, X-ray and radio\footnote{Observations of
A1300 by the MOST at 843 MHz and at higher resolution with the AT
Compact Array at 1.4, 2.4, 4.8 and 8.6 GHz are presented in Paper
III.}.  The properties of the cluster are summarised below.

\begin{itemize}

\item The overall X-ray emission is very extended 
(detected up to $\sim$ 2.8 Mpc from the cluster center in the north)
and is clearly not spherically symmetric.

\item Both the number density and luminosity-weighted galaxy counts 
show good correlation with the principal X-ray extension and
substructure (clump 1), although the latter may not necessarily be at
the same redshift as the cluster.

\item The galaxy velocity dispersion is unusually high (consistent 
with an ICM temperature of $\sim$ 9 keV), and we found some indication
of a temperature gradient in the PSPC data.  The hotter regions, to
the north and east of the cD galaxy, are also those of higher galaxy
density, strengthening the reality of the temperature gradient.


\item The X-ray luminosity is very high ($\sim 1.7\times 
10^{45}$~erg\,s$^{-1}$ in the 0.1--2.4 keV band), supporting a high
ICM temperature $\sim$9 keV on the basis of the empirical $L_X$:$T$
relation (Henry \& Arnaud, 1991).

\item The present X-ray data exclude the presence of a strong cooling 
flow.  However, Henriksen (1993) has found that cooling flows could
form in rich clusters after a merger but before the galaxies have
reached equilibrium.

\item A short HRI exposure of the cluster center (Fig.\ \ref{hri}) 
shows a clear elongated structure near the cluster core, peaking close
to the position of the cD galaxy.

\item Despite its high X-ray luminosity---and correspondingly large 
mass---we did not detect gravitational lensing effects (in the form of
giant arcs or arclets) in the CFHT CCD images. This would suggest that
the mass (luminous and dark) is not centrally peaked (as usually found
in lensing clusters) but rather flat, as implied by the large X-ray
core radius ($\sim$230 kpc), or possibly clumpy.

\item A halo radio source has been discovered close to 
the center of the cluster and located in the same direction as the
inner extension of the X-ray emission (Paper III).

\end{itemize}

At all wavelengths, therefore, it seems as though the results are
consistent and confirm the exceptional nature of A1300.  Many of these
properties may be compared with those found in more easily studied
nearby clusters such as A754 ($z = 0.0541$: Henry \& Briel 1995;
Zabludoff \& Zaritsky 1995) and A2256 ($z = 0.0601$: Briel \& Henry,
1994), which also show substructure and are clearly far from
hydrostatic equilibrium. Signatures of mergers and substructures have
recently been seen as well in ROSAT X-ray data for some distant
clusters, eg. Cl0016+16 (Neumann \& B\"ohringer, 1997) and Cl0939+472
(Schindler \& Wambsganss, 1996).  These signatures were predicted in
N-body/hydrodynamical simulations of hierarchical cluster evolution
(Evrard, 1990a,b; Schindler \& Muller, 1993; Roetigger et al., 1993)
where clusters form through the merger of subclusters. Those
simulations show that when subclusters collide, four effects can be
observed over a time interval $\sim 1 h^{-1}$ Gyr : (i) the X-ray
emission and the galaxy surface density do not coincide; (ii) the
inner region of the ICM is compressed in the direction of the
collision axis and this can appear as a perpendicular elongation in
the X-ray surface brightness distribution; (iii) an anomalously hot
component may be created in the ICM; and (iv) on the large scale an
elongated shape with its axis parallel to the merger axis is
predicted.

In the case of A1300, the first two phenomena predicted by simulations
are absent but we do detect a hot component and an elongated shape.
The high galaxy velocity dispersion, together with the absence of
clear substructure in the velocity histogram, suggests that this
cluster has undergone a merger, but the merging phase may be nearly
over.  In this respect, A1300 resembles the cluster A2142 ($z = 0.09$)
which displays a similar galaxy velocity histogram and velocity
dispersion (Oegerle et al., 1995), integrated gas temperature (from
GINGA data, White et al., 1994), and elliptical X-ray morphology and
luminosity (Henry \& Briel, 1996). Moreover, a detailed temperature
map of A2142 (Henry \& Briel, 1996) provides firm evidence for a
non-relaxed ICM.

A1300 is significant as we appear to be witnessing the end of a merger
at a comparatively early epoch ($z = 0.3$ corresponds to 2/3 of the
age of the universe).  An obvious question to ask is: do all clusters
form at the same epoch or is cluster formation a continuous process?
Comparing A1300 with a similar nearby cluster such as A2142 argues
strongly for the latter hypothesis.  However, an extension of this
work to much higher redshifts will only be possible with the new X-ray
missions such as AXAF to provide the detailed morphology and XMM to
provide spectral information.  Radio observations (Paper III)
reinforce the case for a recent merger in A1300 and highlight the need
for more sensitive high resolution X-ray observations to help
understand the detailed relationship between radio morphologies and
the cluster environment.  Forthcoming ASCA data will provide a much
better constraint on the ICM temperature and any gradients.  Finally,
a weak shear analysis on deeper optical images would be an ideal
complement for mapping the large scale mass distribution and
understanding its dynamical state.

\begin{acknowledgements} It's a pleasure to thank Jean-Luc Starck for his
multi-resolution wavelet analysis program, Emmanuel Bertin for his
source extractor program, and Carlo Izzo for useful information
regarding EXSAS.  RH and AR acknowledge receipt of travel funds
through the Access to Major Research Facilities Program and the
Australia-France Co-operative Fund.

This research has made use of the NASA/IPAC Extragalactic Database 
(NED) which is operated by the Jet Propulsion Laboratory, Caltech, 
under contract with the National Aeronautics and Space 
Administration.
\end{acknowledgements}

\end{document}